\pdfoutput=1

\documentclass[11pt]{article}

\usepackage{authblk}

\usepackage[preprint]{acl}

\usepackage{times}
\usepackage{latexsym}

\usepackage[T1]{fontenc}

\usepackage[utf8]{inputenc}

\usepackage{microtype}

\usepackage{inconsolata}

\usepackage{graphicx}

\usepackage{amsmath}
\usepackage{amssymb}

\usepackage{graphicx}
\usepackage{caption}
\usepackage{subcaption}
\usepackage{lipsum}
\usepackage{mwe}
\usepackage{float} 
\usepackage{resizegather}
\usepackage{hyperref}
\usepackage{enumitem}

\usepackage{xcolor}
\usepackage{soul}
\usepackage{booktabs}
\usepackage{multirow}
\usepackage[normalem]{ulem}
\useunder{\uline}{\ul}{}

\expandafter\def\expandafter\normalsize\expandafter{%
    \normalsize%
    \setlength\abovedisplayskip{3pt}%
    \setlength\belowdisplayskip{3pt}%
    \setlength\abovedisplayshortskip{3pt}%
    \setlength\belowdisplayshortskip{3pt}%
}

\def\equationautorefname~#1\null{(#1)\null}
\def\itemautorefname~#1\null{(#1)\null}
\def\sectionautorefname~#1\null{\S#1\null}
\def\subsectionautorefname~#1\null{\S#1\null}
\def\subsubsectionautorefname~#1\null{\S#1\null}

\title{Context-Aware Language Models for Forecasting Market Impact from Sequences of Financial News}

\author[1,3]{Ross Koval}
\author[2]{Nicholas Andrews}
\author[1]{Xifeng Yan}
\affil[1]{University of California, Santa Barbara}
\affil[2]{Johns Hopkins University} 
\affil[3]{AJO Vista} 
\affil[ ]{\texttt {rkoval@ucsb.edu}}

\begin{document}
\maketitle
\begin{abstract}
Financial news plays a critical role in the information diffusion process in financial markets and is a known driver of stock prices. However, the information in each news article is not necessarily self-contained, often requiring a broader understanding of the historical news coverage for accurate interpretation. Further, identifying and incorporating the most relevant contextual information presents significant challenges. In this work, we explore the value of historical context in the ability of large language models to understand the market impact of financial news. We find that historical context provides a consistent and significant improvement in performance across methods and time horizons. To this end, we propose an efficient and effective contextualization method that uses a large LM to process the main article, while a small LM encodes the historical context into concise summary embeddings that are then aligned with the large model's representation space. We explore the behavior of the model through multiple qualitative and quantitative interpretability tests and reveal insights into the value of contextualization. Finally, we demonstrate that the value of historical context in model predictions has real-world applications, translating to substantial improvements in simulated investment performance.

\end{abstract}
\section{Introduction}
\begin{figure}[t]
\centering
\scalebox{0.95}{
\includegraphics[width=0.5\textwidth]{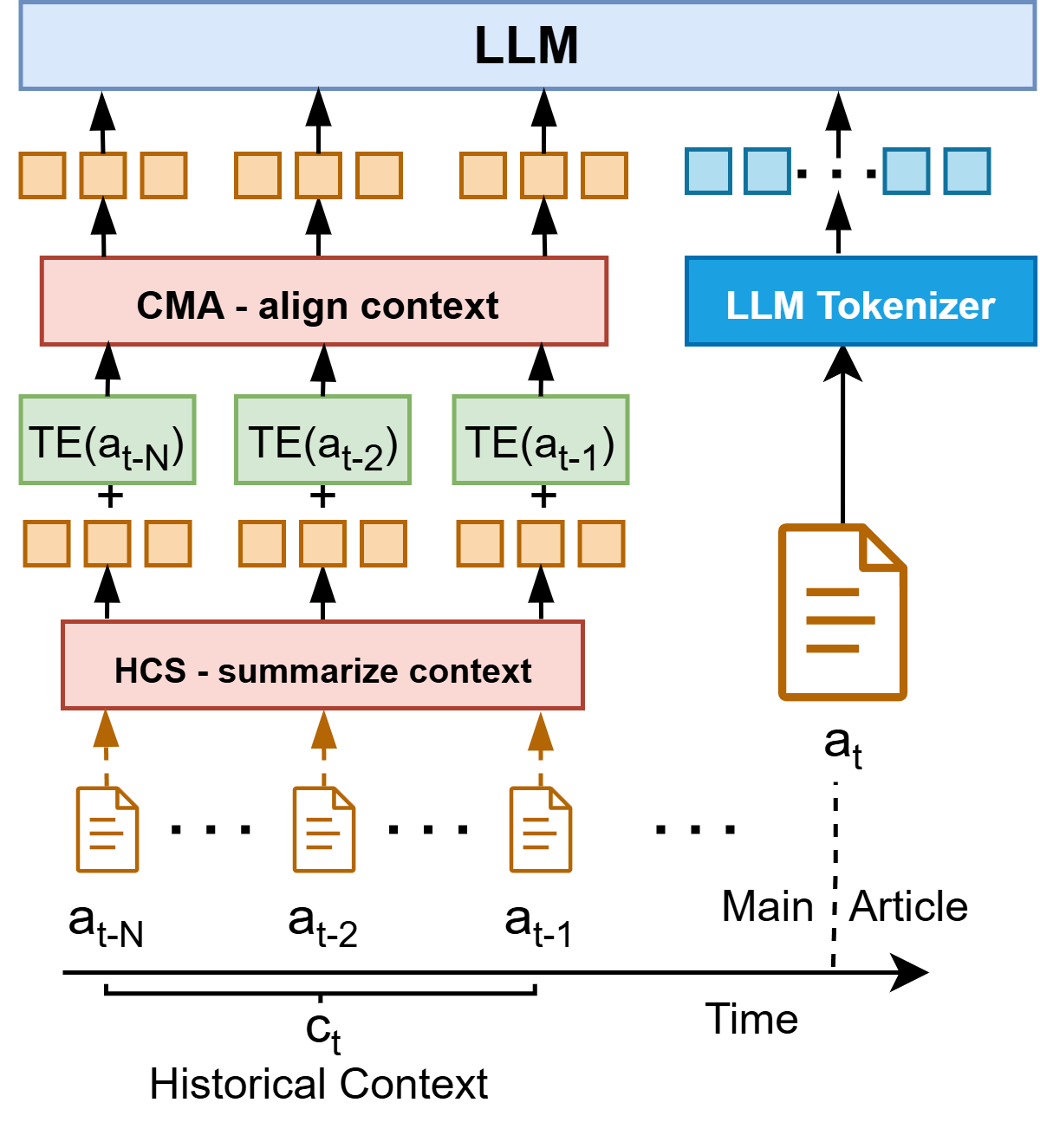}
}
\caption{Overview of our contextualized news forecasting task and proposed Prefix Summary Context (PSC), which efficiently and effectively summarizes historical context with the Historical Context Summarizer (HCS) and aligns it in the latent space of a large LM with Cross-Model Alignment (CMA) and Context-Aware Language Modeling (CALM). The resulting PSC embeddings are prepended to the main article tokens, allowing the LLM to contextualize the main article with historical background.}
\label{fig:psc_model_diagram}
\end{figure}
Financial news plays a critical role in the information diffusion process in financial markets and is a well-known driver of stock prices. These news articles cover a variety of events about companies, including their earnings reports, product launches, legal investigations, and changes in corporate structure. These events can have a substantial impact on company business operations and stock prices. However, it is difficult for investors to be able to identify which articles will materially impact markets because it not only requires being able to understand the sentiment of the articles, but also the more challenging task of recognizing the novelty of the information reported and gathering sufficient historical background to interpret this information in the appropriate context. We hypothesize that the most rigorous test of understanding financial news is being able to predict the market reaction. A complete understanding requires both a local understanding of the information contained in the article but also the broader context in which that news occurs, including prior coverage of related events, as news events often develop over long periods of time.

However, it is not clear \textit{a priori} the best way to retrieve the most relevant context for a given news article and how to optimally incorporate that context into the processing of that article. While the context length of language models has increased significantly, simply concatenating all of the articles together is not necessarily the most efficient or effective method. Firstly, the near-quadratic time and space complexity of the computation makes long contexts computationally challenging.
Secondly, language models have struggled to effectively understand long documents, particularly if irrelevant context is included in their context windows, as they have been shown to exhibit positional biases and ignore content lost in the middle \citep{liu2023lost, mohtashami2024random, li2024long}.

In this work, we investigate how to efficiently and effectively incorporate historical context into language models to better interpret financial news, and demonstrate that gains from our approach lead to significant improvements in financial forecasting and trading performance. In summary, we make the following key contributions:
\begin{enumerate}[leftmargin=*]
  \item We systematically evaluate how historical context improves language model understanding of financial news, demonstrating consistent gains across contextualization methods and time horizons (\autoref{tab:main_results}, \autoref{tab:num_contexts}).
  \item We propose an efficient and effective contextualization method that demonstrates state-of-the-art performance on this challenging task compared to a comprehensive set of baselines (\autoref{sec:methods}, \autoref{tab:main_results}).
  \item We introduce a hybrid retrieval method that captures both the relevance and timeliness of historical context, enhancing the effectiveness of contextualization (\autoref{sec:retrieval}, \autoref{tab:retrieval}).
  \item We probe the model through multiple interpretability methods to reveal insights into the value of historical context and its significant impact on investment applications (\autoref{sec:staleness}).
\end{enumerate}
\paragraph{Broader Impact} We hope this work will inspire future research in modeling the temporal sequence of textual documents and using small LMs to efficiently provide large LMs with more context, particularly as the context length and in-context learning capabilities of LLMs continues to grow, as our findings suggest that retrieval-augmented long contexts underperform forms of context compression and summarization. To support this research direction, we release the code at: \url{https://github.com/rosskoval/calm_fin_news}.

\section{Related Work}

\subsection{Long Context Modeling}
With the advances in retrieval augmented generation and in-context learning \citep{guu2020retrieval, dong2022survey}, there is great interest in extending the context lengths of language models \citep{dai2019transformer, beltagy2020longformer, Zaheer2020, guo2022longt5, jiang2023mistral}. However, given the computational complexity of long contexts, recent work has focused on compressing them into more efficient representations \citep{lester2021power, ge2023context, chevalier2023adapting} that can be leveraged by language models. These methods reduce the time complexity of self-attention at inference time, but often require expensive pretraining procedures to train the base model as a summary compressor. Other approaches use small or frozen LMs to more efficiently encode the context \citep{yen2024long, monteiro2024xc, tan2024lloco}, but lack any form of length compression, and tend to rely on multiple stages of pretraining to learn separate cross-attention weights. 

\subsection{Financial Prediction}
The impact of financial news on stock prices has been extensively studied in the academic literature. \citet{tetlock2007giving, tetlock2011all, fedyk2023can, briere2023stock} find that investors respond to news sentiment in the short-term but the magnitude and duration of the reaction depends upon the novelty of the news content. \citet{chen2022expected, xie2023pixiu, lopez2023can} study the ability of language models to predict stock prices from financial news, finding that performance generally improves with model size. \citet{hu2018listening, xu2018stock, sawhney2021fast, sawhney2021hyperbolic, Agarwal-etal-2022-hyphen, ang2022guided} propose methods to use multimodal streaming news and social media data to make high-frequency predictions of stock prices. 
 
\section{Problem Statement}\label{sec:problem}

\paragraph{Task Formulation}\label{sec:formulation}
Following \citet{xu2018stock, chen2022expected}, we adopt the task of predicting the direction of the stock price $P_t$ change over the course of short-term and long-term horizons (days) $h\in \{\mathrm{7D}, \mathrm{30D}\}$ following the publication of the news article $a_t$ at time $t$.
$$ Y_{t} = \mathrm{Sign}(P_{t+h} - P_{t}) $$
We refer to $a_t$ as the main article and primary input, which we wish to contextualize. To enhance the understanding of the main article, we include $N$ relevant historical articles about the same company for context:
$$c_t = \{a_{t-N},..., a_{t-1}\}$$
For computational efficiency, we select the $N=5$ most recent articles about the same company as the historical context in our main experiments (\autoref{tab:main_results}), but investigate different methods to select the most relevant historical context in \autoref{sec:retrieval} and increase the number of historical contexts in \autoref{tab:num_contexts}. 

We evaluate the performance on this binary classification task with AUC because the continuous scores are more informative than their discrete predicted classes for investment management applications.
\paragraph{Data Acquisition and Curation}
We collect \textit{stock-specific} financial news articles in English from \href{https://www.factset.com/marketplace/catalog/product/streetaccount}{FactSet StreetAccount} for US-based publicly-traded companies. These articles cover a variety of events and news sources. We follow the filtering criterion in \citet{chen2022expected} for data quality, detailed in \autoref{sec:appendix:data_curation}. Our sample contains more than 3,000 public companies in the US, encompassing a diverse range of firm sizes and industries.
\paragraph{Data Statistics and Task Formulation}
We temporally partition the data into training (2010-2014), validation (2015-2016), and test (2017-2023) sets. We provide summary statistics in \autoref{tab:stats}.
\begin{table}[htp]
\centering
{\small
\scalebox{0.90}{%
\begin{tabular}{@{}lrrr@{}}
\toprule
\multicolumn{1}{c}{{\ul }}                                                     & \textbf{Train} & \textbf{Validation} & \textbf{Test} \\ \midrule
Start Date                                                                     & Jan-2010         & Jan-2015              & Jan-2017     \\ \midrule
End Date                                                                       & Dec-2014         & Dec-2016              & Dec-2023     \\ \midrule
\# Samples                                                                     & 129,146           & 46,931                & 149,409        \\ \midrule
\begin{tabular}[c]{@{}l@{}} \# Companies \end{tabular}                 & 2,997          & 2,944               & 3,618         \\ \midrule
\begin{tabular}[c]{@{}l@{}} $\mathrm{TE_1}$ \end{tabular}                 & 15         & 14              & 12          \\ \midrule
\begin{tabular}[c]{@{}l@{}} $\mathrm{TE_5}$ \end{tabular}                 & 115        & 112              & 110
\\ \bottomrule
\end{tabular}
}
\caption{Summary Statistics of the characteristics of news articles based on the average values in each sample split. $\mathrm{TE_i}$ indicates the average time elapsed in number of days between the publication dates of the main article and the $i$th most recent historical context article.}
\label{tab:stats}
}
\end{table}
\section{Proposed Method}\label{sec:methods}
\paragraph{Prefix Summary Context (PSC)}\label{sec:psc_method}
We propose an efficient and effective method to incorporate historical context into a language model's understanding of the main article $a_t$, which we denote as Prefix Summary Context (PSC). Our method adapts a small language model encoder, the Historical Context Summarizer ($\mathrm{HCS}$), to encode and summarize the historical context articles $c_t$. The main article $a_t$ is processed with a large language model ($\mathrm{LLM}$), and the summarized historical context is incorporated via learned prefix embeddings prepended to the input embeddings of the main article.

More specifically, we adapt a small language model encoder to serve as a historical context summarizer ($\mathrm{HCS}$). We train $\mathrm{HCS}$ to learn $M$ summary embeddings per historical news article of length $L$ by inserting learnable summary tokens after every $L//M$ tokens, intended to aggregate information in multiple dimensions:
\[
a_{t-i} = [\,s_1,\, x_1, \dots, x_{k},\, s_2,\, \dots,\, x_{L},\, s_M\,]
\]
where \( s_1, \dots, s_M \) are learnable summary tokens interleaved within the input token sequence \( x_1, \dots, x_L \) of historical context $a_{t-i}$ at regular intervals. We extract the contextualized hidden states of these summary tokens as dense summary embeddings $\mathrm{SE}$ of the input:
$$\mathrm{SE(a_{t-i})=[HCS(s_1 ),..,HCS(s_M)]\in \mathbb{R}^{M \times d_{CE}}}$$
We expect these summary embeddings to provide multiple perspectives of the same article \citep{chang2022multi} that can be selectively attended to depending upon the content of the main article. Dense embeddings have been found to convey almost as much information as the original text itself \citep{morris2023text}, so we expect that learning them end-to-end to provide compact yet expressive representations of the original text.

To help the model distinguish the temporal relevance of each context article, we add time-based embeddings $\mathrm{TE}$, based on the distance between article publication dates. We concatenate the time-enhanced summary embeddings together to form the Summary Context ($\mathrm{SC}$) of $P_L = N \times M$ summary embeddings. 
$$\mathrm{SC_t=[SE(a_{t-N})+TE(a_{t-N} ),…] \in \mathbb{R}^{ P_L \times d_{CE}}}$$

Prior to concatenation, we set the position index of all the prefix summary embeddings to be 0 and then the first token index of the main article to be 1. This is because most large LMs use rotary position embeddings \citep{su2024roformer} in the self-attention computation, and the prefix summary embeddings, containing summarized embeddings from multiple articles, should not be treated equivalently to main article tokens. We allow the time-based embeddings to distinguish the timeliness of articles from each other. This design choice allows context length extrapolation in which more historical context articles can be provided at inference time than during training \citep{chen2023extending, jin2024llm}. 
\paragraph{Cross-Model Alignment} We introduce a Cross-Model Alignment (CMA) module to learn to align the summary context (SC) representations from the historical context summarizer into the token embedding space of the large LM using multiheaded attention (MHA).

Let $E_{\text{vocab}} \in \mathbb{R}^{|V| \times d_{\text{LLM}}}$ be the token embedding matrix of the large LM, where $|V|$ is the vocabulary size and $d_{\text{LLM}}$ is the embedding dimension. We compute the Prefix Summary Context (PSC) using multihead attention with query $Q = \mathrm{SC} \cdot W^Q$, and keys and values $K = V = E_{\text{vocab}}$, where $W^Q \in \mathbb{R}^{d_{\text{CE}} \times d_{\text{LLM}}}$ projects the context summarizer's summary embeddings to the dimensionality of the LM's token embeddings:
\[
\mathrm{PSC} = \mathrm{MHA}(Q, K, V) \in \mathbb{R}^{P_L \times d_{\text{LLM}}}
\]

This cross-attention mechanism allows each summary embedding in SC to attend over the entire vocabulary of the large LM, producing aligned representations that lie in the span of the large model's token embeddings. This step can be viewed as an alignment between the semantic spaces of the two language models, a concept inspired by alignment methods in other domains, such as vision \citep{liu2024visual, li2023blip} and time series \citep{jin2023time, zhou2023one, sun2023test, liu2024taming}. This design ensures that the Summary Context is compatible in the input embedding space of the large LM, and in doing so, imposes the inductive bias that PSC can be explained by a learned weighted combination of the token embeddings of the large LM.

We prepend PSC to the token embeddings of the most recent article, which are passed to the large LM's transformer layers for contextualization: 
$$\mathrm{CE_t}=\mathrm{LLM}( [\mathrm{PSC_t}; \mathrm{E}_{\text{vocab}}[a_t]])$$
where $\mathrm{E}_{\text{vocab}}[a_t]$ represents the token embeddings for the main article $a_t$.

Overall, our design provides two key advantages: 
\begin{enumerate}
  \item It exploits the power of a large LM on the main article, which is most important to the prediction.
  \item It enables efficient incorporation of more historical articles as context with a small LM that summarizes the key information.
\end{enumerate}
Further, we note that our method introduces a minimal number of trainable parameters and allows the historical context to interact with each other through the LLM's transformer layers, contrary to interleaved cross-attention layers \citep{alayrac2022flamingo,monteiro2024xc,yen2024long} which introduces significantly more parameters. We conduct ablations of our architecture design in \autoref{sec:ablations}, which demonstrate their value.

\paragraph{Context-Aligned Language Modeling (CALM)}
To enable the historical context summarizer to distill and encode relevant information that enhances the LLM’s comprehension of the main article, we pretrain the context summarizer ($\mathrm{HCS}$) and alignment module ($\mathrm{AM}$) with context-aligned language modeling (CALM), a pretraining objective inspired by retrieval-augmented language modeling \citep{guu2020retrieval}.
In this step, we learn to use PSC to predict the tokens in the main article. Given the encoded historical context $\mathrm{PSC_t}$, main article $a_t$, and token index $i$, the loss function is:
$$\mathrm{L_{CALM} = -\sum_{i} \log p(a_{i,t} | PSC_t; a_{1,t},...,a_{i-1,t})}$$
This objective encourages the context summarizer and alignment module to learn to encode contextual information that is predictive of future news text, thereby improving the LLM's ability to interpret the market impact of the main article. 
\paragraph{Implementation Details}\label{sec:imp_details}
We initialize \textbf{Mistral-7B} \citep{jiang2023mistral} as the large LM and \textbf{DeBERTa-base} \citep{he2021debertav3} as the historical context summarizer. During pretraining, we freeze the large LM and update only the context summarizer and alignment module. During supervised finetuning, we use LoRA \citep{hu2021lora} to finetune the large LM, while fully training all newly introduced parameters. Please see \autoref{sec:appendix:implementation} for more details.

\begin{table*}[t]
\centering
{\small
\scalebox{.85}{%
\begin{tabular}{@{}llllcc@{}}
\toprule
\textbf{Model Class} & \textbf{Method} & \textbf{\# Params} & \textbf{\# Contexts} & \textbf{7D} & \textbf{30D} \\ \midrule

\multirow{8}{*}{\textit{Zero-Shot LLMs}} 
  & LMD-Sent \citep{loughran2011liability} & 0 & 0 & 50.71 & 51.32 \\
  & FinBERT-Sent \citep{araci2019finbert} & 110M & 0 & 51.19 & 51.38 \\
  & FinMA-7B, Prompting \citep{xie2023pixiu} & 7B & 0 & 51.37 & 51.64 \\
  & FinMA-7B, Prompting \citep{xie2023pixiu} & 7B & 5 & 51.49 & 52.18 \\
  & Llama3-8B, Prompting \citep{lopez2023can} & 8B & 0 & 51.22 & 51.47 \\
  & Llama3-8B, Prompting \citep{lopez2023can} & 8B & 5 & 51.68 & 51.94 \\
  & Llama3-70B, Prompting \citep{lopez2023can} & 70B & 0 & 52.76 & 53.35 \\
  & Llama3-70B, Prompting \citep{lopez2023can} & 70B & 5 & 53.31 & 53.81 \\ \midrule

\multirow{3}{*}{\textit{Financial Forecasting}} 
  & HAN \citep{hu2018listening} & 20M & 5 & 53.17 & 54.32 \\ 
  & FAST \citep{sawhney2021fast} & 110M & 5 & 53.62 & 54.75 \\ 
  & HYPHEN \citep{Agarwal-etal-2022-hyphen} & 110M & 5 & 55.38 & 56.37 \\ \midrule

\textit{Single-Article Baseline} 
  & SINGLE & 7B & 0 & 55.73 [0.07] & 56.50 [0.06] \\ \midrule

\multirow{5}{*}{\textit{Long-Context Baselines}} 
  & CONCAT-FULL & 7B & 5 & 56.43 [0.09] & 57.52 [0.08] \\
  & CONCAT-PREFIX & 7B & 5 & 56.90 [0.09] & 57.74** [0.09] \\ 
  & MDS \citep{caciularu2023peek} - SUM + CONCAT & 7B & 5 & 56.62 [0.09] &  57.61 [0.08] \\
  & MDS \citep{caciularu2023peek} - CONCAT + SUM & 7B & 5 & 56.47 [0.09] & 57.22 [0.07] \\ 
  & HIERARCHICAL \citep{ivgi2023efficient} & 7B & 5 & 56.95* [0.13] & 57.66 [0.12] \\ \midrule
\multirow{1}{*}{\textbf{Proposed}} 
  & \textbf{PSC} & 7B & 5 & \textbf{58.24*} [0.09] & \textbf{59.12**} [0.08] \\
\bottomrule
\end{tabular}
}}
\caption{Main Results (higher is better): Model performance on the test set of our contextualized news prediction task at different forecasting horizons. All models after and including "SINGLE" use Mistral-7B as the base model. All results indicate the AUC of the model's predicted probabilities reported in percentage units. "[]" indicate the standard deviation of the results for that method over 3 training runs with different random seeds. *, ** indicate that the performance of our proposed model is statistically better ($p<0.01$) than that of the highest-performing baseline according to the Wilcoxon Signed-Rank Test. The last row (bolded) indicates our proposed method PSC.}
\label{tab:main_results}
\end{table*}

\section{Baselines}\label{sec:baselines}
We explore a comprehensive set of baseline methods for modeling long contexts, including single-article and multi-article prediction. For all experiments in \autoref{sec:methods} and \autoref{tab:main_results}, single-article predictions are made solely based on the most recent article, while multi-article prediction incorporates historical context based on the previous $N$ articles about the same company. We explore more sophisticated methods to retrieve the most relevant context in \autoref{sec:retrieval}.

\subsection{Zero-Shot Baselines}\label{sec:single_article}
First, we establish some zero-shot, pretrained baselines to highlight the difficulty of the task. We begin with the Loughran McDonald financial sentiment dictionary (\textbf{LMD-Sent}) and the pretrained financial sentiment classifier \textbf{FinBERT-Sent} \citep{araci2019finbert}. We also include leading open-source LLMs: Llama3-Instruct-8B, Llama3-Instruct-70B, \textbf{Llama3, Prompting} \citep{touvron2023llama}, as well a financial instruction-tuned model \textbf{FinMA} \citep{xie2023pixiu}. For each model, we prompt it to predict the direction of the market reaction, according to the prompt developed in \citet{lopez2023can}. We explore multiple variations of this prompt, detailed in \autoref{sec:appendix:prompting}, and report the one with the best validation performance. The consistently poor performance of these baselines and their inability to effectively leverage historical context highlights the challenges LLMs face in interpreting the impact of complex factors driving market reactions.  
\subsection{Financial Forecasting Baselines}\label{sec:domain}
We include financial domain-specialized baseline methods, including HAN \citep{hu2018listening}, FAST \citep{sawhney2021fast}, and HYPHEN \citep{Agarwal-etal-2022-hyphen}, which were designed to make stock movement predictions from sequences of financial text, and currently represent the state-of-the-art on this type of financial forecasting task. 

\subsection{Long-Context Baselines}
We evaluate several strong long-context modeling baselines to assess the value of incorporating historical context. For all experiments, we use Mistral-7B \citep{jiang2023mistral}, a 7B-parameter model with an 8K context window, to ensure fair comparison against our proposed method. As a primary baseline, we finetune Mistral-7B on single news articles (\textbf{SINGLE}) to isolate the effect of adding historical context.

\paragraph{Concatenation}\label{sec:concat} The most direct contextualization method is to simply concatenate the previous articles and the current article together, denoted as \textbf{CONCAT-FULL}, allowing the self-attention mechanism to interact them. We also include a version \textbf{CONCAT-PREFIX} that concatenates the first paragraph of each historical context article. We sort them in ascending order of time and delimit each article with special tokens. Since the context length of most LLMs exceeds 8K tokens, this approach can easily accommodate more than 10 articles. In both cases, we use Mistral-7B to encode this long context and finetune on the task. This represents the most direct approach of leveraging long-context capabilities without specialized contextualization. We extract the representation of the main article tokens with mean pooling, now contextualized with historical articles.

\paragraph{Multi-Document-Summarization} MDS involves the summarization of a set of related documents, which is a natural approach to extract the most relevant information from the historical context. We use QAMDEN \citep{caciularu2023peek}, a state-of-the-art MDS model shown to outperform GPT-4 \citep{caciularu2023peek}. We consider two strategies. First, we summarize the historical context and prepend this summary to the main article, denoted as \textbf{MDS-SUM+CONCAT}. Alternatively, we concatenate the historical context and the main article together, and summarize them jointly, denoted as \textbf{MDS-CONCAT+SUM}. We pass the resulting output to \textbf{Mistral-7B}, which is then finetuned on the task as in \autoref{sec:single_article}. 

\paragraph{Hierarchical Encoding} Alternatively, a hierarchical encoding approach can be used \citep{yang2020beyond, zhang2019hibert, izacard2020leveraging, ivgi2023efficient, bertsch2024unlimiformer}. In this approach, adapted from \citet{ivgi2023efficient} and denoted as \textbf{HIERARCHICAL}, each article is encoded independently, and the resulting token representations are concatenated and globally contextualized across articles by a transformer encoder, further detailed in \autoref{sec:appendix:hierarchical}.

\section{Experimental Results and Analysis}
For all experimental results, we would like to highlight the large size of the test set (150K samples), allowing us to detect most small differences in model performance with statistical significance. \textbf{We demonstrate the substantial practical value of significant but seemingly modest absolute gains in classifier performance in \autoref{sec:port_sims}}.

\paragraph{Historical Context} The main results in \autoref{tab:main_results} highlight the challenging nature of the task, but \textbf{we find broad consistency in the relative performance of contextualization methods across time horizons}.

Overall, we find that all forecasting horizons benefit consistently and significantly from the inclusion of historical context. Contrary to \citet{chen2022expected}, we do not find a significant degradation in model performance for longer time horizons, which we hypothesize is due to our contextualization method. From a practical perspective, we highlight that longer-term financial prediction is typically more useful for investors since it reduces transaction costs, so these gains are likely to be even more valuable in real-world applications, which we confirm in \autoref{sec:port_sims} with portfolio simulations. 

\paragraph{Context Length} In \autoref{tab:num_contexts}, we compute the forecasting performance and language model (cross-entropy) loss as a function of the number of historical context articles with the PSC contextualization method. In this experiment, we only train the model with 5 context articles but test with up to 20 since our method allows for context length extrapolation. 

In this result, we view the LM loss as an approximation of LM's understanding of the main article \citep{salazar2019masked, du2024understanding}. The results provide evidence towards our hypothesis that model performance on our task demonstrates language understanding as we find that both forecasting performance \textit{and} LM loss improves with the number of contexts incorporated. 

Moreover, we find that the method performance continues to improve up to $N=15$ context articles, but the gains diminish after 10. We note that the method's ability to generalize to more articles at inference time is particularly valuable as the number of news articles over a given period of time varies considerably with the size and industry of the company. Furthermore, as shown in \autoref{tab:stats}, the average age of context articles $c_{t-N}|N\ge5$ is more than 3 months, which supports our hypothesis that, despite the age, the context is helpful in interpreting the recent article, as the market is clearly no longer reacting to the old context article.

\paragraph{Contextualization Method}
While we find that all contextualization methods provide a material benefit over their corresponding single-article baselines, we find that approaches that summarize the relevant context into a more concise and informative form, either in the language or the latent space, tend provide better performance than simple concatenation, which is the most computationally expensive. 

We conjecture that the observed trend is due to several factors. Firstly, concatenation does not impose the inductive bias that the most recent article is the most critical to the prediction. Secondly, concatenation is likely the most sensitive contextualization method to positional bias and irrelevant context. Recent work has shown that even SOTA commercial LLMs exhibit positional bias in the information that they pay attention to, irrespective of its relevance, and can become confused by irrelevant context \citep{liu2023lost, mohtashami2024random, li2024long, hsieh2024ruler}. Since the most relevant historical context may not be optimally positioned within the context window, or may only be partially relevant to the main article, we suspect that these challenges are exacerbated on a task with such a low signal-to-noise ratio. 

\begin{table}[htp]
\centering
{\small
\scalebox{0.90}{%
\begin{tabular}{@{}cccc@{}}
\toprule
\textbf{\# Contexts} & \textbf{LM Loss} & \textbf{7D} & \textbf{30D} \\ \midrule
0            & 2.10    &  55.73     &    56.50              \\
1            & 1.95    &  56.60     &    57.47              \\
2            & 1.84    &  57.48     &    58.39              \\
5            & 1.80    &  58.24     &    59.12              \\
10            & 1.75    &  58.59     &    59.44              \\
\textbf{15}            & \textbf{1.72}    &  \textbf{58.66}     &    59.50              \\ 
\textbf{20}            & 1.73    &  58.64     &    \textbf{59.52}              \\ 
\bottomrule
\end{tabular}
}
\caption{Results indicate the forecasting performance and LM loss of the PSC contextualized model while varying the number of the most recent articles to include as context.}
\label{tab:num_contexts}
}
\end{table}
\paragraph{Retrieval Method}\label{sec:retrieval}
In the experiments so far, we assume that the $N=5$ most recent historical articles about the same company provide the best context for the main article (\textbf{Time}). However, the most recent articles may not always be the most relevant context as news stories can develop over longer time periods.

We test this hypothesis by instead retrieving the $N=5$ most relevant articles about the same company according to semantic similarity. To do this, we experiment with three strong but generic retrievers: \textbf{SBERT} \citep{reimers2019sentence}, \textbf{Contriever} \citep{izacard2021unsupervised}, \textbf{InstructOR} \citep{su2022one}. Additionally, we create our own domain-specific, text-encoder (\textbf{FinSim}) pretrained on the training data with the DiffCSE framework \citep{chuang2022diffcse}. We further incorporate time into the measure by applying an exponential decay with a half-life of 6 months to the similarity score to capture both context similarity and timeliness (\textbf{TimeFinSim}). Given main article $a$, historical context $c$, time elapsed $t$, and half-life $H$: 
$$\mathrm{TimeFinSim(a, c, t) = FinSim(a, c) \cdot e^{-\frac{\ln(2)}{H} t}}$$ 

For each target article, we retrieve the $N=5$ closest articles from the same company within the last year and pass them to PSC for contextualization in temporal order. We include more details about the retrievers in \autoref{sec:appendix:retrieval}.

As shown in \autoref{tab:retrieval}, we find that semantic similarity-based retrieval is only better than time-based when the retriever is specialized to the financial domain (\textbf{FinSim}) and that a hybrid approach (\textbf{TimeFinSim}) performs significantly better. We conjecture that since the retrieval set is all financial articles about the same company, generic retrievers are not sufficiently discriminative to connect related events together. 
\begin{table}[htp]
\centering
{\small
\scalebox{0.90}{%
\begin{tabular}{@{}cccc@{}}
\toprule
\textbf{\begin{tabular}[c]{@{}c@{}}\end{tabular}} \textbf{Retriever} & \textbf{\# Contexts} & \textbf{7D} & \textbf{30D} \\ \midrule
Time & 5 & 58.24 & 59.12 \\
SBERT & 5 & 58.02 & 58.80 \\
Contriever & 5 & 58.10 & 58.96 \\
InstructOR & 5 & 58.17 & 59.11 \\
FinSim & 5 & 58.81 & 59.74 \\   
\textbf{Proposed-TimeFinSim} & 5 & \textbf{59.12} & \textbf{60.15} \\ \bottomrule    
\end{tabular}
}
\caption{Results indicate the AUC performance of different text retriever models used to retrieve the $N=5$ most relevant articles to contextualize the main article with PSC contextualization method.}
\label{tab:retrieval}
}
\end{table}
\section{Model Analysis and Interpretability}\label{sec:model_analysis}
In this section, we explore the behavior of the model from different perspectives and reveal insights into the value of historical context and potential impact on investment applications. 

\paragraph{Staleness of News}\label{sec:staleness}
Financial news articles do not always report novel information. Stale news may cover information that has already been priced into the market, in which case it can have a different market impact. To better understand the impact of stale news on our results and the value of contextualization, we must first introduce a measure of news staleness. For this purpose, we define article staleness to be the average similarity of the article to the preceding $N=5$ articles about the same company, using SBERT \citep{reimers2019sentence} for document embeddings $\mathrm{E}$:
$$\mathrm{staleness}(a_t) = \frac{1}{5} \sum_{i=1}^{N=5} \, \mathrm{sim}(E(a_t), E(a_{t-i}))$$
This metric will assign higher staleness scores to articles which overlap substantially with previous articles about the same company, and estimate how much novel information is reported. We categorize the test data into 3 levels of staleness based upon this measure. 

\begin{figure}[!t]
\centering
{\small
\scalebox{0.85}{
\begin{tabular}{|l|l|}
\hline
\multicolumn{1}{|c|}{\textbf{Historical Context, N=1}} & \multicolumn{1}{c|}{\textbf{Main Article}} \\ \hline
\begin{tabular}[c]{@{}l@{}}Harte Hanks announces\\ sale of Trillium business\\  to Syncsort for \$112m \\ in cash...\hl{Harte Hanks} \\ \hl{will use substantially all}\\ \hl{of the net proceeds from}\\ \hl{the sale to retire its} \\ \hl{outstanding credit facility}\\ ...additional details will \\ be provided at a later \\ date..\end{tabular} & \begin{tabular}[c]{@{}l@{}}Harte-Hanks to delay\\ 10-K filing...\hl{compli-}\\ \hl{cations required to}\\ \hl{account for the sale} \\ \hl{of the company’s} \\ \hl{trillium business}\\ (and calculate the \\ tax therefor),...delays\\ and complications in \\ performing its annual \\ goodwill impairment\\ analysis, and additional\\ time-consuming review, \\ testing and evaluation.\end{tabular} \\ \hline
\end{tabular}
}}
\caption{Sample news that highlights the value of historical context in understanding the recent news article. Without knowing the terms of the sale, filing delays would typically be interpreted negatively. 
However, the sale allows the company to pay off its outstanding debt. We vary the number of historical context articles ($\mathrm{N}$) and show that more context leads to consistently improved language understanding and forecasting performance.}
\label{fig:case_study}
\end{figure}

As shown in \autoref{tab:staleness}, we find that the historical contextualization provides consistent benefit across all articles but has the largest relative improvement for the stalest articles relative to their histories, as the model can learn to adjust its predictions based upon the novelty of the information and its effect within the related context. It is also important to note that high article novelty (low staleness) is likely associated with less relevant contexts, which could be viewed as distracting information and ultimately confuse the model from the main article \citep{liu2023lost}. However, our method still provides value in these instances, suggesting that it is robust to less relevant contexts, which we attribute to its ability to distill relevant information.

\begin{table}[htp]
\centering
{\small
\scalebox{0.90}{%
\begin{tabular}{@{}cccc@{}}
\toprule
\textbf{Staleness Level} & \textbf{Single, N=0} & \textbf{Contextualized, N=5} & \textbf{$\Delta$} \\ \midrule
1                      & 56.94         & 58.31    &  1.37             \\
2                      & 56.51         & 59.37    &  2.86             \\
3                      & 56.39         & 59.72    &  3.33           \\ \bottomrule
\end{tabular}
}
\caption{Results indicate the 30D AUC performance of the single article model baseline and the PSC contextualized model, as well as the difference between them ($\Delta$), across articles with different levels of staleness relative to their histories, reported in percentage points.}
\label{tab:staleness}
}
\end{table}
\paragraph{Case Study and Qualitative Analysis}\label{sec:case_study}
We conduct a qualitative analysis of model predictions to further understand the effects of historical contextualization and its impact on the model behavior. To this end, we sort the difference in model predictions between the single article and historical contextualized model, and examine the sample with the largest difference. In this example, displayed in \autoref{fig:case_study}, the single-article model predicts that the filing delay reported in the most recent article will lead to a negative return, but the additional context of an article from a few months prior reveals the reason is due to the sale of a business unit that will enable the company to pay down its debt and reduce its interest burden. As a result, investors viewed the news favorably and the stock rallied considerably over the following month.

\paragraph{Portfolio Simulations}\label{sec:port_sims}
In \autoref{tab:port_sims}, we demonstrate the economic value of our methodology with portfolio simulations. We form monthly long-short (\textit{market-neutral}) portfolios \citep{fama2015five} by sorting stocks based on the average 30D-horizon model predictions from the past month, detailed in \autoref{sec:appendix:port_sims}. For comparison, we simulate two common trading strategies: Price Momentum \citep{jegadeesh1993returns} and the Fama-French 6-Factor pricing model \citep{fama2018choosing}, described further in \autoref{sec:appendix:ff6}. Following \citet{cong2021alphaportfolio}, we include net performance that includes conservative estimates of the impact of transaction costs on portfolio implementation.
 
The resulting performance of our contextualization method generates investment performance that is economically and statistically better than the single and multi-article baselines. Relative to the single article baseline, we note that a seemingly modest absolute gain in AUC ($\sim$2.5\%) from our approach results in a more than 50\% relative improvement in investment performance. These results demonstrate that the contextualized model provides significant predictive value beyond the single-article baseline in a real-world trading setting. 

\begin{table}[htp]
\centering
{\small
\scalebox{0.80}{%
\begin{tabular}{@{}cccc@{}}
\toprule
\textbf{Method} & \textbf{Net Return} & \textbf{Volatility} & \textbf{Net Sharpe Ratio} \\ \midrule
Price Momentum    & 5.30 & 18.33 & 0.29 \\
Fama-French 6-F    & 6.99 & 14.51 & 0.48 \\ \midrule
SINGLE, $N=0$ & 9.04 & 13.40 & 0.67  \\
CONCAT, $N=5$ & 11.35 & 13.30 & 0.85  \\
HIERARCH, $N=5$ & 12.04 & 13.55 & 0.89  \\ \midrule
\textbf{Proposed-PSC, $N=5$} & \textbf{14.13} & \textbf{13.33} & \textbf{1.06}\\
\textbf{+ TimeFinSim}        & \textbf{15.02} & \textbf{13.25} & \textbf{1.14}\\
\bottomrule
\end{tabular}
}
\caption{Annualized portfolio statistics of simulated investment performance, expressed in percentage units. $N$ denotes the number of historical context articles used. ``Net'' performance includes an estimate of the impact of transaction costs, detailed in \autoref{sec:appendix:port_sims}.}
\label{tab:port_sims}
}
\end{table}

\section{Conclusion}
We explore the value of historical context in the ability of language models to understand the market impact of financial news, and find that it provides a consistent and significant improvement in performance across methods and time horizons. To this end, we propose an efficient and effective contextualization method that allows the use of a large LM on the main article while using a small LM to summarize the historical context. Through multiple interpretability tests, we reveal insights into the value of historical context and demonstrate that it directly translates to improvements in simulated investment performance.

\paragraph{Limitations}
Our experiments demonstrate that historical context significantly improves the ability of the model to predict the market reaction of a news article, and that those improvements translate to gains in simulated investment performance. Additionally, we show that context summarization and compression methods perform better than simple concatenation. However, we acknowledge our experiments focus on open-source LLMs under 100B parameters and that results may differ at larger scales. We also hope to expand this research to non-English news for international companies in the future. We expect that our contextualization method can generalize effectively to multilingual and international markets, particularly since our methods relies on a modular architecture in which the context summarizer and large LM can be independently replaced or adapted to different languages or domains. Indeed, we hypothesize that gains from historical contextualization may be even larger in these less efficient markets or in languages with sparser or noisier news coverage, where each individual article may be shorter, less self-contained, or lack detailed background. In such settings, the ability to distill and align prior news coverage becomes even more important. Finally, while we incorporate conservative estimates of transaction costs in our investment simulations, real-world trading requires more detailed consideration of trade execution and risk management, which we leave to future work. Please note that our financial prediction system is intended for research use and that portfolio results are presented for illustration purposes only, not as investment advice.

\section*{Acknowledgements}
We would like to thank AJO Vista and FactSet for providing access to the data. The authors are solely responsible for the content and views expressed in this publication and do not reflect those of the affiliated institutions.

\bibliography{custom}

\appendix
\section{Appendix}
\label{sec:appendix}

\subsection{Portfolio Simulations}\label{sec:appendix:port_sims}
In \autoref{tab:port_sims}, we demonstrate the economic value of our model predictions using portfolio simulations. In these results, all contextualization methods use \textit{Time} as the retrieval method (i.e. we retrieve the 5 most recent articles about the same company as the historical context), except for the last row which uses our proposed hybrid retrieval method to perform the context article selection. 

First, we filter the investment universe of stocks according to sufficient liquidity requirements to ensure feasibility of portfolio implementation, including a minimum market capitalization of \$250M and daily average value of shares traded of \$1M. 

Then, we form monthly long-short (market-neutral) quintile portfolios according to \citet{fama2015five}. To do so, we sort stocks based on the average 30D model predictions from news articles about companies in the past 1 month. We demean score values by industry averages to remove any potential effect of industry bets or risks. Then our portfolios are formed by buying those in the top 20\% of average scores and shorting those in the bottom 20\% of average scores on a monthly basis in equal proportions. Please note that these portfolios are market-neutral and therefore have essentially no correlation with broad market indices. 

In \autoref{tab:port_sims}, we include conservative estimates of the impact of transactions costs on portfolio implementation. We follow the turnover-based method used in \citet{cong2021alphaportfolio}, which conservatively estimates the annual transaction cost as 0.01 times the annual 1-way portfolio turnover. 

\subsection{Fama-French 6-Factor Model (FF-6)}\label{sec:appendix:ff6}
We use the Fama-French 6-Factor model \citep{fama2018choosing}, which consists of 6 financial factors, including size, beta, value, momentum, profitability, and investment, to make predictions of stock returns by estimating the linear regression model over the 10 years prior to the start of the simulation (Jan 2007 - Dec 2016) using monthly stock returns. We use the estimated model to make forward stock return predictions using these 6 financial factors on a monthly basis. Then, we form long-short quintile portfolios based on the monthly predictions \citep{fama2015five}.  

\subsection{Data Curation}\label{sec:appendix:data_curation}
Since FactSet StreetAccount uses human curation to only provide relevant, market-moving news in the newsfeed, we do not perform any additional steps to filter the data for impactful news. 

Following \citet{chen2022expected}, if the news article was published after market open (9:30 am), then we consider the news article to be published the next trading day. We only include articles originally written in English according to the following criteria \citep{chen2022expected}: they are tagged relevant for only one company; they are longer than 100 characters or shorter than 10,000 characters; they contain less than 10\% of numerical characters; they have less than 90\% Jaccard similarity to a previous article (to remove potential duplicate articles). We require that each main article possess at least 5 historical articles about the same company within the past year as well as available stock returns.

There is a slight class imbalance so we randomly downsample the majority class to ensure balanced class ratios. All models are optimized using Binary Cross-Entropy as the loss function.

\subsection{Pretrained Language Models}
We develop all Transformer-based models in PyTorch and source all pretrained checkpoints from HuggingFace. 

\subsection{Prefix Summary Context}\label{sec:appendix:psc}
In the Prefix Summary Context (PSC) method, we tune the number of attention heads in the alignment module (AM) over \{2, 4, 8, 16\} and the number of summary embeddings per historical context article over \{1, 2, 5, 10\} according to validation set performance. 

\subsection{Architecture Ablations}\label{sec:ablations}
In \autoref{tab:ablations}, we ablate the choice of the alignment module between the representation space of the context summarizer and large LM, including both a \textbf{Linear} projection and also a fully connected \textbf{MLP} network. We find that the use of the cross-attention mechanism in the alignment module to be important for convergence, particularly for direct finetuning without CALM pretraining. We conjecture that this effectively forces the resulting attention output to be within the convex hull spanned by the large model's token embeddings, and suspect that this makes them more likely to be in-distribution and thus receive attention weight. We expect this improvement to be more pronounced when using common parameter efficient finetuning methods that do not update the LMs token embeddings, not allowing them to adapt distributions to the new prefix. The challenges of direct finetuning from continuous prompts \citep{li2021prefix} and different modality spaces \citep{yen2024long, ge2023context} have been confirmed in the literature, which consequently use a staged training approach.

\begin{table}[htp]
\centering
{\small
\scalebox{0.90}{%
\begin{tabular}{@{}cc@{}}
\toprule
\textbf{Alignment Method} & \textbf{30D AUC} \\ \midrule
Linear                    & 57.50                 \\
MLP                       & 57.37                 \\
Proposed-CMA                       & 58.33             \\ \midrule
Linear + CALM             & 58.29                 \\
MLP + CALM                & 58.67                 \\
Proposed-CMA + CALM                & 59.12             \\ \bottomrule
\end{tabular}
}
\caption{Results indicate the 30D AUC performance of different alignment methods between the output of the context summarizer and the input embeddings of the large LM with and without CALM pretraining.}
\label{tab:ablations}
}
\end{table}

\subsection{Pretrained Baselines}
We include pretrained financial sentiment classifier \textbf{FinBERT-Sent + Linear} \citep{araci2019finbert} applied at the sentence-level \citep{alanis2022benchmarking}:
\begin{equation*}
\resizebox{0.48\textwidth}{!}{%
     $
     \text{FinBERT-Sent} = \frac{\text{\#PositiveSentences - \#NegativeSentences}}{\text{\#TotalSentences}}
     $
 }
 \end{equation*}

To be consistent across all text embedding models, we apply mean pooling at the token-level to aggregate contextualized token representations to the sequence level.

For some baseline models in which additional inputs or modalities are incorporated, such as proprietary sentiment scores or stock price time series, we do not include these additional inputs (and we do not have access to them) in the model in order to present a fair comparison of model types and isolate the effects of historical contextualization.

\subsection{Prompting}\label{sec:appendix:prompting}
For the zero-shot Llama-3-7B and Llama-3-70B baselines, we have explored a variety of prompting strategies, listed below. We do not find much variation in the performance of each, but find (1) developed by \citet{lopez2023can}, to perform best according to the validation set and report those results in \autoref{tab:main_results}. 

\begin{enumerate}[label=(\arabic*),leftmargin=*]

\item
\begin{quote}\small\ttfamily
Forget all your previous instructions. Pretend you are a financial expert. You are a financial expert with stock recommendation experience. Answer "YES" if good news, "NO" if bad news in the first line. Then elaborate with one short and concise sentence on the next line. Is this news article good or bad for the stock price of \{company name\} in the next \{time horizon\} days? Article: \{news article\}
\end{quote}

\item
\begin{quote}\small\ttfamily
Classify the following financial news article(s) as 'Positive' if it will lead to a positive stock return in the next \{time horizon\} days and 'Negative' if it will lead to a negative stock return in the following \{time horizon\} days. Please only respond with 'Positive' or 'Negative': \{news article\}.
\end{quote}

\item
\begin{quote}\small\ttfamily
Predict if the following financial news article is 'Positive' if it will cause a positive stock return in the next \{time horizon\} days and 'Negative' if it will cause a negative stock return in the following \{time horizon\} days. Please only respond with 'Positive' or 'Negative': \{news article\}.
\end{quote}

\item
\begin{quote}\small\ttfamily
You are a financial analyst with expertise in market reactions. Based on the following news article(s), predict the likelihood of a positive stock return for \{company name\} in the next \{time horizon\} days. Respond with a probability between 0\% and 100\%. Article: \{news article\}.
\end{quote}

\end{enumerate}

For each prompt response (except for \#4), we extract the softmax probabilities for the two corresponding positive and negative tokens and re-normalize to arrive at a continuous score. If historical context is provided, then the historical news articles are prepended to the main article. 

\subsection{Statistical Significance}\label{sec:stats}
In \autoref{tab:main_results}, we report the standard deviation of results across 3 different training runs with different random seeds. The variability of results from the random seeds results from the randomness in the training process caused by random initialization of the classification layer weights and the random order of training samples during stochastic gradient descent optimization. We also report the results from the Wilcoxon Signed-Rank test, which is a pairwise test of statistical significance conducted with bootstrapped samples from the test set.

\subsection{Hierarchical Encoding}\label{sec:appendix:hierarchical}
For the hierarchical approach, we select a randomly-initialized, causal Transformer to globally contextualize the local (article-level) token representations. We apply mean pooling across the token embeddings of the most recent article $a_t$. We tune the number of layers $\in \{2, 4, 8\}$ and the number of attentions head $\in \{2, 4, 8\}$ according to validation set performance.

\subsection{Retrieval Models}\label{sec:appendix:retrieval}
For Instructor embedding models \citep{su2022one} for context retrieval, we use the following prompt: "Represent the financial news article for retrieval: \{news article\}". We use the \href{https://huggingface.co/hkunlp/instructor-base}{instructor-base} to ensure similar parameter counts across retrieval models. 

For FinSim, we conduct the DiffCSE-Domain pretraining process over the training data for a maximum of 25K training steps or until the loss on the validation set increases, using the same hyperparameter configuration and settings as \citet{chuang2022diffcse}. We initialize the model from the pretrained checkpoint of DeBERTa-base and additionally use that as the fixed generator (masked language model) model. We tune the tradeoff between the contrastive and replaced token detection (RTD) objective over $\{0.10, 0.25, 0.50, 0.75, 0.90\}$ according to validation set performance. Please see \citet{chuang2022diffcse} for more details on the framework. 

For TimeFinSim, we use the time distance from the historical news article date to the main article date to apply an exponential decay (half-life = 180 days) to the FinSim similarity score to capture both the similarity and timeliness of the historical context. Given main article $a$, historical context $c$, time elapsed $t$, and half-life $H$: 
$$\mathrm{TimeFinSim(a, c, t) = FinSim(a, c) \cdot e^{-\frac{\ln(2)}{H} t}}$$ 

\subsection{Training Details and Hyperparameter Tuning}\label{sec:appendix:implementation}
We perform all experiments on a single Tesla A100 GPU with 80G in memory. We use AdamW to optimize all parameters. For all finetuned models, we use an effective batch size of 32 with gradient accumulation. We train all models for up to 5 epochs with early stopping based off validation set performance for test evaluation. 

For long-context models based on Mistral-7B, we finetune with low-rank adapters (LoRA) \citep{hu2021lora}. We tune the learning rate over \{1e-6, 3e-6, 5e-6, 7e-6, 1e-5\} and LoRA rank parameter over $r\in \{16, 32, 64\}$ according to validation set performance. We apply LoRA adapters to all linear layers of the base model. 

In all main baselines, based on Mistral-7B, we apply mean pooling across the hidden states of the main article tokens, which is then passed to a classification layer to make a binary prediction.

For computational constraints, we train all models using mixed precision training and gradient checkpointing to satisfy GPU memory constraints, and clip gradient norms. For DeBERTa-based models, we finetune in FP16 precision, while we use BF16 for Mistral-based models. For LoRA-based finetuning, we always set the value of the alpha parameter to be equal to double the value of rank parameter. 

\end{document}